\documentclass[twocolumn,prb,groupedaddress,preprintnumbers,amsmath,amsfonts,amssymb,floatfix,showpacs]
{revtex4}

\usepackage{graphicx}% Include figure files
\usepackage{dcolumn}% Align table columns on decimal point
\usepackage{bm}% bold math
\usepackage{epsfig}
\usepackage[usenames,dvips]{color}
\usepackage[english]{babel}
\usepackage{multirow}

\begin{document}
\title{Electronic and magnetic properties of the Ti$_5$O$_9$ Magn\'{e}li phase}

\author{I.~Slipukhina}
\email[Corresponding author: ]{i.slipukhina@fz-juelich.de}
\author{M.~Le\v{z}ai\'{c}}
\affiliation{{Peter Gr\"unberg Institut and Institute for Advanced Simulation, Forschungszentrum J\"ulich and JARA,  D-52425 J\"ulich, Germany}}

\date{\today}

\begin{abstract}
Structural, electronic and magnetic properties of Ti$_5$O$_9$ have been studied by \textit{ab initio} methods in low-, intermediate- and high-temperature phases. We have found the charge and orbital order in all three phases to be non-stable, and the formation of Ti$^{3+}$-Ti$^{3+}$ bipolaronic states less likely as compared to Ti$_4$O$_7$. Several quasidegenerate magnetic configurations were calculated to have different width of the band gap, suggesting that the reordering of the unpaired spins at Ti$^{3+}$ ions might at least partially be responsible for the changes in conductivity of this material.

\end{abstract}


\maketitle

\section{Introduction}
\label{sec:introduction}

For several decades binary transition-metal oxides and related compounds
have attracted increasing attention as resistive-switching materials. These are materials which can be switched between a high- and a low-resistance states and are very promising candidates for the next generation non-volatile resistive random access memory (RRAM) \cite{}. The resistive switching in many of these technologically relevant oxides is claimed to be based on the formation and disruption of highly-conductive filaments, through which the current flow is realized \cite{}. However, still very little is known about the composition, structure and dimensions of these filaments.

\begin{figure}[htbp]
	\begin{center}
\includegraphics[width=1.0\hsize]{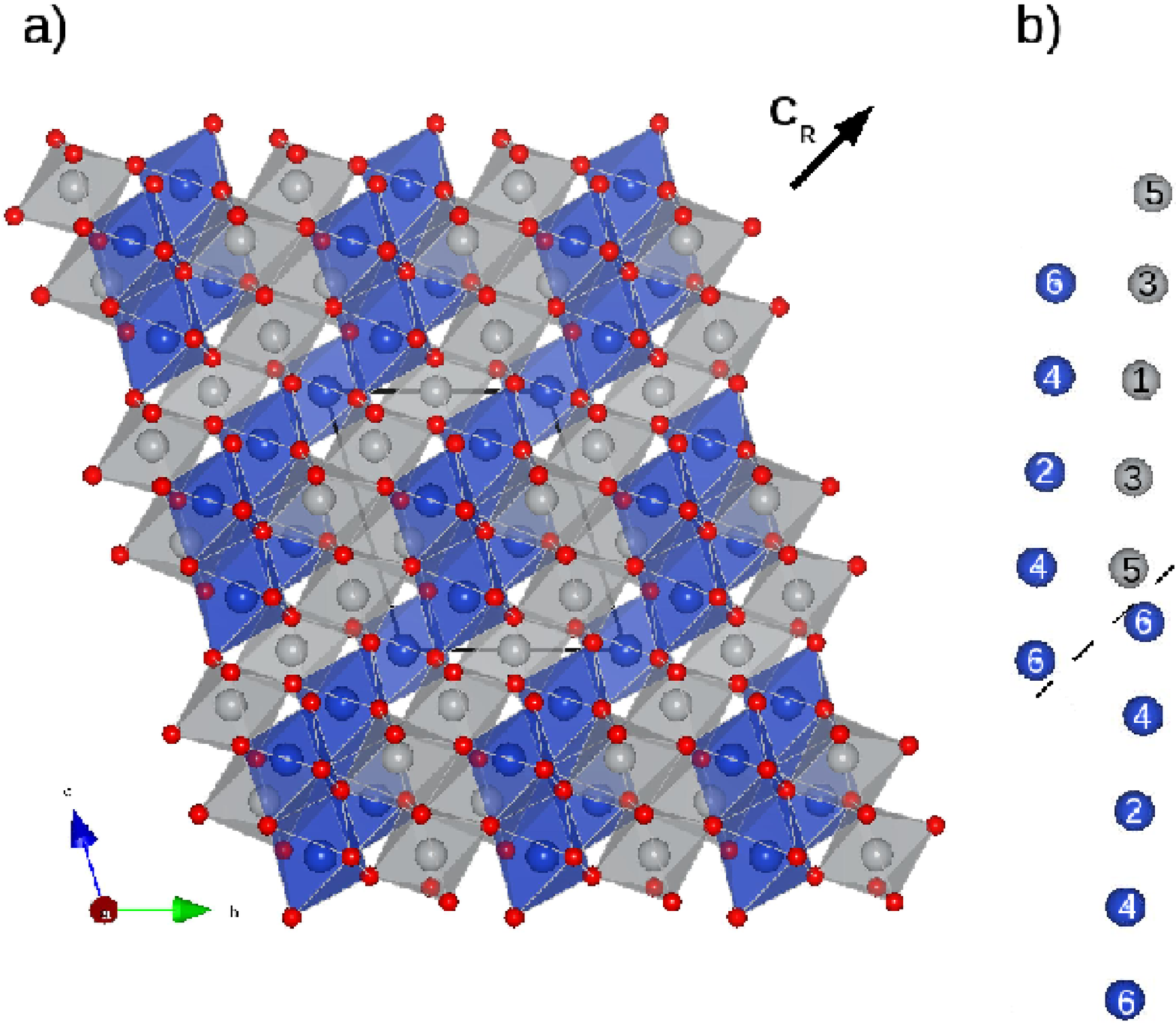}
\includegraphics[width=0.6\hsize]{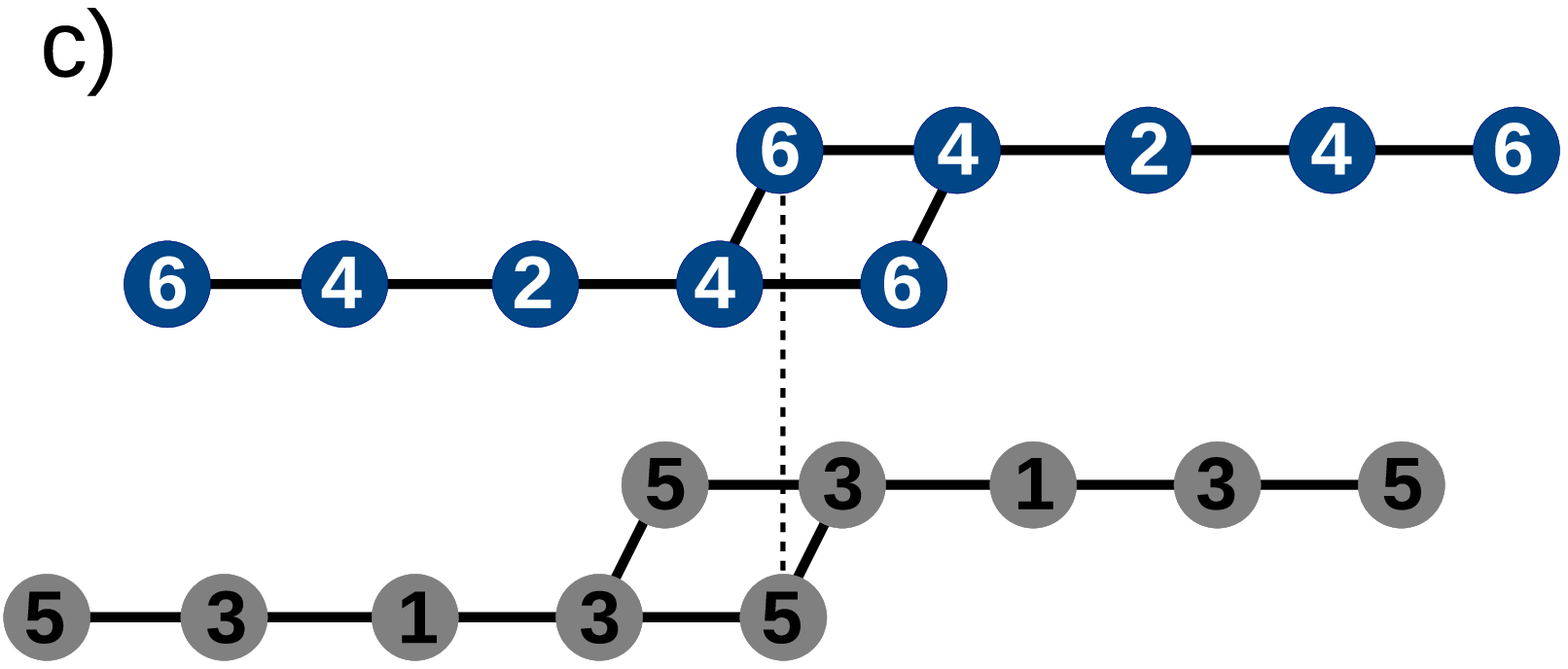}
 		\caption{(color online) The crystal structure of the Ti$_5$O$_9$: spatial view (a),  schematic view of the chain structure (b) and two adjacent Ti chains (c). The Ti atoms in the adjacent chains, which run parallel to the pseudorutile c$_\mathrm{R}$ axis, are shown in grey and blue, while oxygen atoms are shown in red. The black dashed line indicates the shear plane. \label{fig:Fig_1}}
	\end{center}
\end{figure}

Recent low-temperature conductivity and in situ current-voltage measurements confirmed that the conducting filaments in Pt/TiO$_2$/Pt \cite{Kwon_2010}, as well as Fe-doped SrTiO$_3$ \cite{Waser_2010} are composed of Ti$_4$O$_7$ and Ti$_5$O$_9$ Magn\'{e}li phases and their mixtures. These compounds belong to the homologous series Ti$_{n}$O$_{2n-1}$ ($3\leq n\leq9$) of Ti oxides with TiO$_2$ and Ti$_2$O$_3$ as the end members.  Structurally they are related with rutile: there are adjacent rutile-like chains of $n$ edge-sharing TiO$_6$ octahedra, running parallel to the pseudorutile $c_R$ axis (see Fig.\ref{fig:Fig_1}); the octahedra at the end of each of the chains share their faces with the next pseudorutile chain. With the exception of Ti$_2$O$_3$, these compounds posses a triclinic structure
and they are mixed-valence compounds with two Ti$^{3+}$ (3$d^1$ electronic configuration) and ($n-2$) Ti$^{4+}$ ($3d^0$) ions per formula unit (in an ionic picture). Several members of this homologous series, as well as the isostructural vanadium Magn\'{e}li phases V$_n$O$_{2n-1}$, are known to exhibit phase transitions with temperature, at which their conductivity changes drastically (it is the largest for $n=3,4$ and 5 and reduces for the members with higher $n$). These transitions were interpreted in terms of the localization of the electrons in the cation sublattice. Indeed, the presence of both Ti$^{3+}$ and Ti$^{4+}$ ions provides several possibilities of charge distribution at cation sites, resulting in various charge-ordered states. 
 
Take Ti$_4$O$_7$ for example, which undergoes two consecutive phase transitions with temperature: a semiconductor-semiconductor transition at 130~K and a semiconductor-metal transition at 150~K \cite{Marezio_1972}. X-Ray diffraction studies \cite{Marezio_1973} on Ti$_4$O$_7$ single-crystals revealed that at low temperatures Ti$^{3+}$ ions are covalently bonded to form the so-called \textit{bipolarons} -- singlet electron pairs (Ti$^{3+}$-Ti$^{3+}$ pairs). In the low-temperature (LT) semiconducting phase these pairs order so that they occupy alternate pseudorutile chains. In the high-temperaure (HT) phase the pairing is believed to be absent and all the Ti-cations are in the same oxidation state Ti$^{3.5+}$, resulting in further increase of the conductivity and metallic behavior. The nature of the intermediate temperature (IT) phase remains rather unclear, however it is commonly believed that the (semiconductor-semiconductor) transition from LT to IT phase is accompanied with the reordering of the Ti$^{3+}$-Ti$^{3+}$ pairs \cite{Marezio_1973}. It was assumed that in the IT phase bipolaros distribute in Ti-chains with a long-range order that requires a fivefold superstructure (bipolaron liquid state \cite{Lakkis_1976}). Later, the structure of the IT phase was revisited and suggested to be more complicated with the presence of bipolarons and a long-range order of Ti valences \cite{Le_Page_1984}. Recent photoemission experiments \cite{Taguchi_2010} interpreted the IT phase in Ti$_4$O$_7$ as a highly anomalous state sandwitched between the mixed-valent Fermi liquid (HT phase) and charge ordered Mott-insulating phase (LT phase).

Two distinct phase transitions were also observed in Ti$_5$O$_9$, at 128~K and 139~K, accompanied by the conductivity decrease by a factor of 3 (Ref.~\onlinecite{Marezio_1977}) (or 50 according to Bartolomew and Frankl \cite{Bartolomew_1969}). While LT and IT phases are found to be semiconducting, the conductivity in the HT phase increases with increasing temperature, which means that this phase is not a true metal and its nature is not clear. Compared to Ti$_4$O$_7$, for which a stable charge-ordered state was shown to exist at low temperature, the charge order in Ti$_5$O$_9$ is rather uncertain, since the number of Ti ions in the pseudorutile chains is odd, preventing the formation of the Ti$^{3+}$ long-range ordered pairs \cite{}. X-ray studies performed by Marezio \textit{et al} \cite{Marezio_1977}, did not reveal any localization of charges and/or formation of covalent bonds between the Ti$^{3+}$ cations in Ti$_5$O$_9$. However, in order to explain the semiconducting properties of the LT and IT phases, which in Ti$_4$O$_7$ are related with the formation of Ti$^{3+}$-Ti$^{3+}$ pairs, they suggested a microdomain structure of the material with Ti$^{3+}$ mostly present at domain walls, making the detection of the charge localization and Ti$^{3+}$-Ti$^{3+}$ bonds formation by x-ray techniques rather impossible. The observed EPR signal the authors associated with the presence of unpaired Ti$^{3+}$. Watanabe \cite{Watanabe_2009} has recently studied the structural changes across the phase transitions in Magn\'eli phases by means of Raman spectroscopy. He observed two semicondutor-semiconductor transitions, at 137~K and 130~K, and revealed a fine structure of the LT phase, somewhat resembling the IT phase of Ti$_4$O$_7$. He suggested a considerable disorder for the LT phase of Ti$_5$O$_9$ and a rather itinerant nature  of the 3$d$ electrons in the IT and HT phases as in the HT phase of Ti$_4$O$_7$. The author did not observe any clear tendency as to the nature of the charge ordering in the considered Magn\'eli phases and concluded that in the compounds with odd $n$ like Ti$_5$O$_9$ the charge order is unstable.

The effect of pressure on the phase transitions in Ti$_n$O$_{2n-1}$ ($n$=4, 5, 6) Magn\'eli phases was studied in Ref.~\onlinecite{Ueda_2002}. 
In contrast to a commonly accepted bipolaronic picture of the metal-insulator transition in these compounds, which is attributed to the strong electron-lattice coupling, the authors proposed their own explanations of the physics behind the phase transitions. The remarkably enhanced pressure effect on the phase transition temperatures as $n$ goes from 6 to 4 the authors explained by a delicate interplay between the electron-lattice coupling and electron correlations. In this respect charge ordering in Ti$_4$O$_7$ is explained to be due to a dominant electron correlations with a subsequent bipolarons formation by an electron-lattice coupling. Ti$_6$O$_{11}$, on the contrary, is believed to be a small polaron insulator due to a prevailing electron-lattice coupling.  
Ti$_5$O$_9$ is suggested to be in between these two extreme cases. The formation of small polarons, localized on one atomic site and randomly distributed over the lattice, explains rather well the unclear charge distribution in higher $n$ Magn\'eli phases, observed in earlier experimental works \cite{Marezio_1977,Le_Page_1983}. The substantial Curie contribution, observed in Ti$_5$O$_9$ at low temperatures \cite{Mulay_1970}, which can be associated with the presence of upaired localized moments, is also in line with the proposed small polaron picture. A distinct pressure dependence of the resistivity in the HT phase of Ti$_5$O$_9$ allowed the authors to assume that the system is just at the crossover region from large to small polarons. Such crossover in the systems with intermediate electron-coupling is often associated with a metal-insulator transition.

As for the magnetic properties, Ti$_5$O$_9$, as well as Ti$_4$O$_7$ and Ti$_6$O$_{11}$ phases show antiferromagnetic (AFM) behavior with N\'{e}el temperature ($T_N$) of about 130~K \cite{Keys_66}. %For these compounds the anomalous behavior of the magnetic susceptibility was observed below the $T_N$, whic was suggested to be due to phase transformation or trapping of the delocalized electrons.
According to magnetic susceptibility measurements in Ti$_5$O$_9$, there are two transitions which occur at 128 and 138 K, in contrast to Ti$_4$O$_7$ with only one transition in magnetic susceptibility \cite{}. The $T_N$ is very close to the phase transition temperature, which allows one to relate the magnetic ordering with the phase transitions observed in this compound.  
At low temperatures ($T<40$~K) the measurements show a Curie behavior, while for the 40~K$<T<$128~K the temperature-independent susceptibility might be due to Van Vleck paramagnetism. Susceptibility in the HT phase is clearly temperature independent. 
Based on the measured molar susceptibility in Ti oxides, the effective magnetic moment for Ti$_5$O$_9$ was calculated to be of 0.245~$\mu_{\mathrm{B}}$ and it was concluded that a relatively small percentage of the unpaired electrons (7.4\%) contributes to the Curie-Weiss behavior, while a large percentage (92.6\%) contributes to its metallic (free electron gas) behavior; the small Weiss constant indicates that these spins do not interact strongly \cite{Danley_1972}.

Although rather well studied experimentally, there is very little known about the mechanism behind the transitions in Magn\'{e}li phases theoretically. Up to today, there are only a few DFT calculations on Ti$_4$O$_7$ \cite{Eyert_2004,Leonov_2006,Liborio_2009,Weissmann_2011,Padilha_2014}, and to our knowledge, they are completely lacking for higher-$n$ phases. Therefore, in this work we aimed to develop a fundamental understanding of the mechanisms that underlie the phase transitions in Magn\'{e}li phases like Ti$_5$O$_9$, which, due to its high electrical conductivity and chemical stability, has a promising application not only in electrochemical engineering, but even in medicine as a neural stimulation electrode \cite{Canillas_2013}. With this aim we performed DFT calculations in order to illuminate the changes in its electronic structure on a microscopic level and establish relations between the structural, electronic and magnetic properties and their implications for conductive phenomena. We believe that the obtained results will be helpfull in interpreting the physics of resistively switching Ti oxides. 

\section{Computational details}
\label{sec:Method}

The electronic and magnetic properties of the different phases of Ti$_5$O$_9$ were studied within the DFT, using the Vienna \textit{ab initio} Simulation Package (VASP) \cite{Kresse1_1996,Kresse2_1996} with projector augmented potentials (PAW) \cite{Blochl_1994,Kresse_1999}. Kinetic energy cut-off of 450 eV and $6\times6\times6$ $\textbf{k}$-points mesh 
was used for the unit cell of 28 atoms. The exchange correlation functional was treated within the generalized gradient approximation (GGA) \cite{Perdew_1996,Perdew_1997}. 
GGA+$U$ approximation in Dudarev's approach \cite{Dudarev_1998} with $U$ applied to $3d$ states of Ti was used to take into account the electronic correlations. 

The structural data for all three phases have been taken from the Ref.~\onlinecite{Marezio_1977}. The structure optimization was performed only for the LT phase with the aim to find the ground state of the system under study. For the sake of simplicity, we do not consider the spin-orbit coupling in our calculations, and we restrict ourselves to the collinear spins only.

\section{Results and discussion}
\subsection{Correlation effects}
\label{sec:Correlation_effects}

It is known that oxygen deficient rutile TiO$_{2-x}$ can be transformed into a Magn\'eli phase, when oxygen nonstoichiometry is sufficiently high ($x>0.001$ \cite{Bursil_1969}). At such concentrations, oxygen vacancies arrange into extended defects, resulting in the shear plane structure (see Fig.\ref{fig:Fig_1}(c)). As a result, the defect states appear in the band gap close to the conduction band minimum, which are occupied by the electrons of Ti$^{3+}$ ions. Numerous studies on TiO$_{2-x}$, including recent hybrid functional calculations \cite{Janotti_2010}, have reported on the effect of the electronic correlations on the position of these defect states and the distribution of Ti$^{3+}$ ions. For Ti$_4$O$_7$ or Ti$_5$O$_9$ with similar peculiarities of the electronic structure, $+U$ approach or hybrid functionals were used to correctly account for the correlation effects and were able to reproduce the ground state properties of the phases \cite{Eyert_2004,Leonov_2006,Liborio_2009,Weissmann_2011,Padilha_2014}. However, the correlations might be less effective in the phases with larger $n$ (i.e., with smaller density of 3$d$ electrons), as it was assumed in Ref. {\onlinecite{Watanabe_2009}} based on the observed lack of the systematic trend in charge ordering in different Magn\'eli phases.

\begin{figure}[htbp]
	\begin{center}
		\includegraphics[width=1.0\hsize]{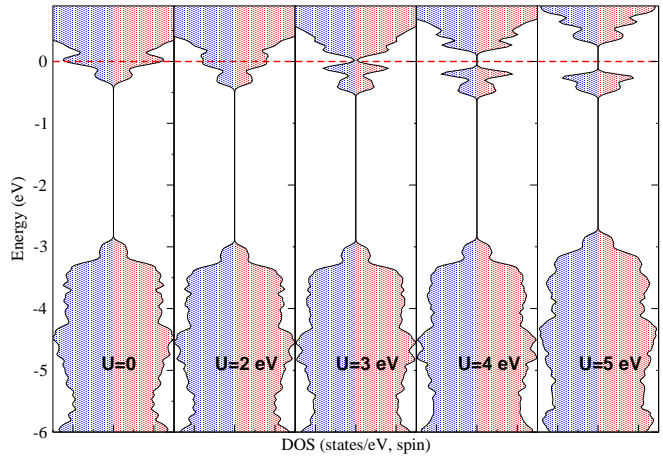}
 		\caption{(color online) The GGA+$U$ spin-polarized DOS, calculated for the LT phase of Ti$_5$O$_9$ at different $U$ for the AFM spin allignment. The Fermi level is at zero energy (shown by red dashed line). Blue and red shading denotes the majority and minority spins, correspondingly.  \label{fig:Fig_2}}
	\end{center}
\end{figure}

In order to ilucidate the importance of electron correlations for the charge localization in Ti$_5$O$_9$, we have peformed calculations within GGA+$U$ approach, assuming the unit-cell model for the order/disorder of the Ti$^{3+}$ and Ti$^{4+}$ ions. To cover the limits of weak and strong electron interactions, the on-site Coulomb energy parameter was varied within the range $0<U<6$~eV. The evolution of the density of electronic states (DOS) with $U$ in the LT phase with the experimental structure is presented in Fig.\ref{fig:Fig_2}; these calculations were performed assuming an AFM spin coupling with a zero total magnetic moment per cell. 

\begin{figure}[htbp]
	\begin{center}
\includegraphics[height=0.6\hsize]{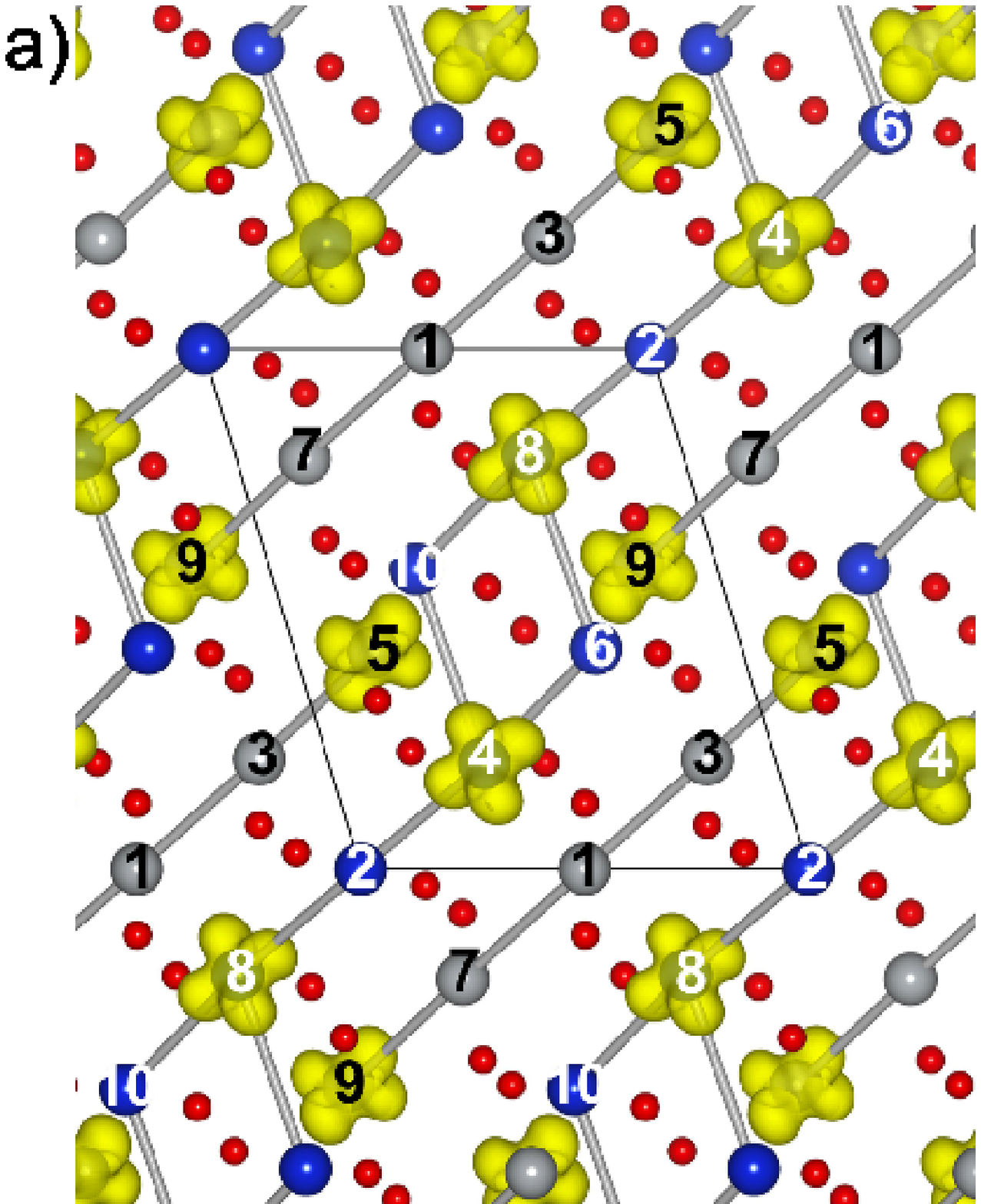}
\includegraphics[height=0.6\hsize]{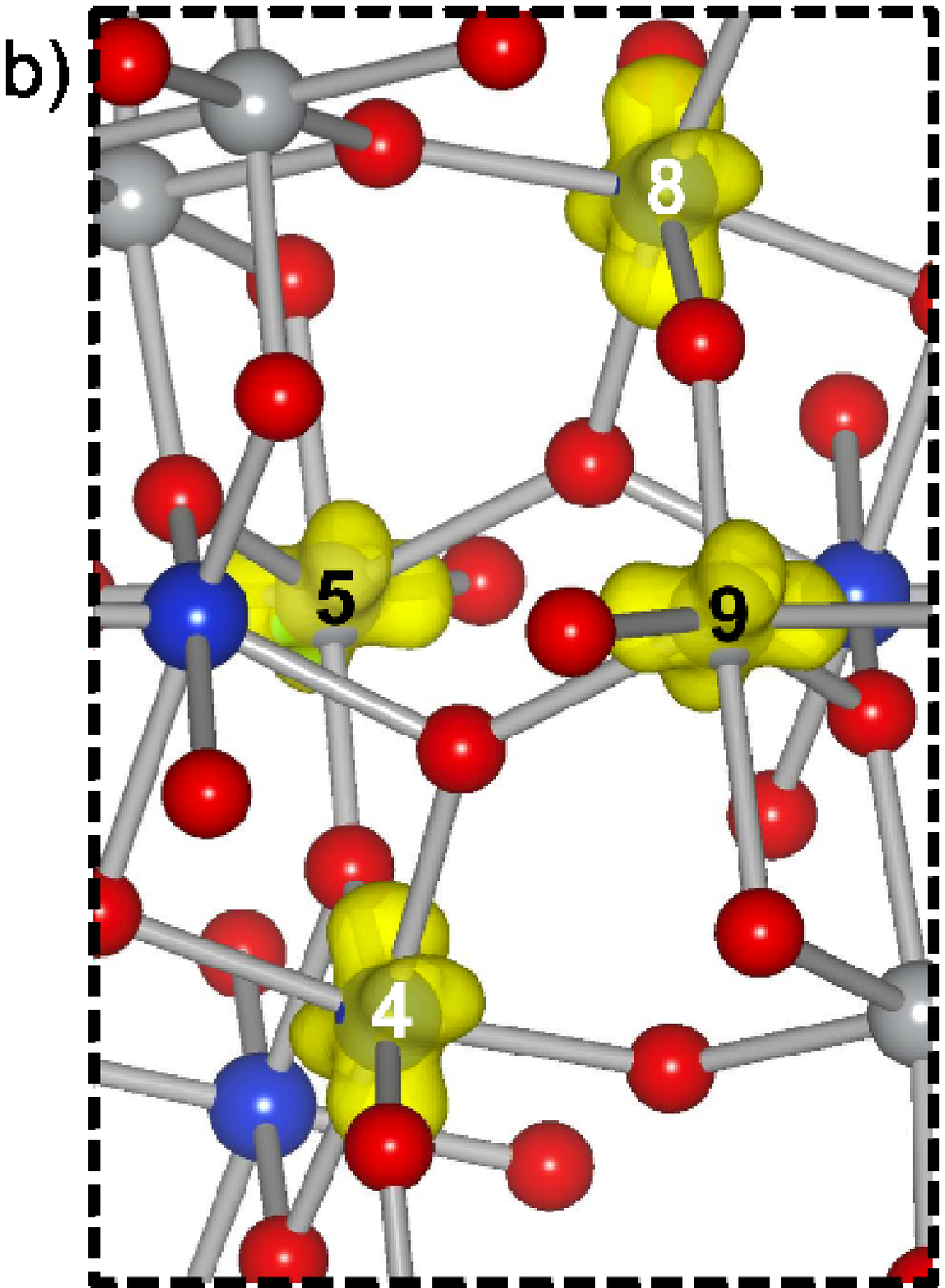}	
\caption{(color online) $3d$ valence charge distribution in the LT phase of Ti$_5$O$_9$: (a) general view with Ti-Ti bonds; (b) zoomed view with Ti-O bonds. The calculated charge density isosurfaces correspond to the Ti$^{3+}$ gap states. The presented results were obtained at $U=6$~eV for the AFM spin configuration with $M_{\mathrm{tot}}=0$. Blue and grey colors distinguish between the Ti atoms in the adjacent pseudorutile chains. Note different orientation of the $t_{2g}$-like orbitals.\label{fig:Fig_3}}
	\end{center}
\end{figure}

As follows from Fig.\ref{fig:Fig_2}, for $U\leq3$~eV the DOS remains metallic with Ti $3d$ states merged into the conduction band. A small band gap of about 0.1~eV between the occupied and unoccupied $3d$ states of Ti opens in both spin channels at $U>3$~eV, and it increases up to 0.7~eV at $U=6$~eV (see Tab.\ref{tab:table_1}). It should be noted that an energy gap of 0.7~eV was also obtained for Ti$_4$O$_7$ at $U=0.4$~Ry (5.4~eV) in Ref.\onlinecite{Weissmann_2011} (the choice of this value of $U$ was justified by comparing the total energy difference between the three phases at different $U$ with the transition temperatures). The topology of the insulating DOS very much resembles that of the Ti$_4$O$_7$ \cite{Leonov_2006,Liborio_2009,Weissmann_2011}. The well-localized states just below the Fermi level are formed by the $3d$ orbitals of Ti possessing non-zero magnetic moment.
Figure~\ref{fig:Fig_3} shows the spatial distribution of these localized gap states, calculated at $U=6$~eV. It is clear that the system is not only charge ordered, but also orbitally-ordered: occupied orbitals at different Ti$^{3+}$ sites are of $d_{xy}$ and $d_{xz}$ (or $d_{yz}$) symmetry, if we define them in the local frame of the octahedra. 
%However, it is difficult to define explicitly the symmetry of each of these orbitals, since the distortions of the corresponding TiO$_6$ octahedra are quite complicated as follows from Tab.\ref{tab:table_2}.
While the conduction band is mainly contributed by Ti $3d$ states, the valence band originates from the oxygen $2p$ states (with a small admixture of Ti $3d$ states due to Ti-O bonds formation) and it is separated from the gap states by an energy interval, which is much larger than the fundamental gap. This gap decreases with the increasing $U$ from around 2.8~eV at $U=0$ to 1.9~eV at $U=6$~eV. The width of the split-off $3d$-band is decreasing as the electron localization increases with the increased $U$. Following the evolution of the DOS as a function of $U$ parameter, it is difficult to make certain conclusions on the appropriate value of $U$ in the case of Ti$_5$O$_9$, since the existing experimental data about the electronic band structure of Ti Magn\'eli phases is quite diverse. According to Mulay \textit{et al} \cite{Mulay_1970}, who studied the cooperative magnetic transitions in titanium oxides, the band gap of Ti$_5$O$_9$ in the LT phase is found to be of 0.035 eV. The authors also report on the gap of 0.041~eV for Ti$_4$O$_7$, while optical measurements in Ref.~\onlinecite{Kaplan_1977} gave a band gap of 0.25~eV; later photoemission spectroscopy experiments revealed a gap of only 0.1~eV\cite{Taguchi_2010}. One is clear: even if the electron-electron interactions are less pronounced in Ti$_5$O$_9$ than in the other Magn\'eli phases with higher concentration of $3d$ electrons, they play an important role in the stabilization of its insulating ground state.

\begin{table}
 \caption{The GGA+$U$ absolute values of magnetic moments $\lvert M \rvert$ (in $\mu_B$) at Ti sites in the LT phase of Ti$_5$O$_9$, calculated for different $U$ (at $J=1$ eV) for the AFM/FM spin allignment. The energy gaps $E_g$ (in eV) and the total energy differences $\Delta{E}=E_{AFM}-E_{FM}$ (in meV/f.u.) are aslo listed.}\label{tab:table_1}
\begin{ruledtabular}
\begin{tabular}{c|c|c|c|c|c}
      & $U$=2~eV& $U$=3~eV &  $U$=4~eV & $U$=5~eV & $U$=6~eV \\
  \hline
    Ti(1) &  0/0.21   &  0/0.23 &  0/0.15    & 0/0.14    &  0/0    \\
    Ti(2) &  0/0.47   &  0/0.52 &  0/0.50    & 0/0.55    &  0/0.44    \\ 
    Ti(3) &  0.29/0.30   &  0.37/0.43 &  0.38/0.45    &  0.28/0.44   &  0/0.56    \\ 
    Ti(4) &  0.19/0.30   &  0.43/0.36 &  0.58/0.43    &  0.71/0.37   &  0.81/0.43    \\
    Ti(5) &  0.27/0.19   &  0.32/0.47 &  0.37/0.49    &  0.53/0.61   &  0.77/0.59    \\ 
    Ti(6) &  0/0.09   &  0/0.26 &  0/0.22    & 0/0.17    &  0/0.12    \\
\hline
    $E_g$  &   0/0       &  0/0        &  0.1/0      & 0.4/0        & 0.7/0 \\
\hline
    $\Delta{E}$  &   55        &  47        &  23      & $-19$        & $-72$ \\
    \end{tabular}
\end{ruledtabular}
\end{table}

\subsection{Structural effects}
\label{sec:Structural_effects}

\begin{table}
 \caption{Ti-O interatomic distances (in \AA) in the LT (115~K) and HT( 295~K) phases of Ti$_5$O$_9$ according to structural parameters from Ref.\onlinecite{Marezio_1977} }\label{tab:table_2}
\begin{ruledtabular}
\begin{tabular}{c c c c c c}
      & 115~K & 295~K & & 115~K & 295~K \\
  \hline
    Ti(1)-O(2) & 1.949    & 1.955  & Ti(4)-O(5)    &  1.978  &  1.987   \\
    Ti(1)-O(3) & 1.963    & 1.970  & Ti(4)-O(6)    &  2.085  &  2.095   \\ 
    Ti(1)-O(4) & 2.009    & 2.015  & Ti(4)-O(9)    &  2.086  &  2.092  \\ 
    Ti(2)-O(1) & 2.013    & 2.008  & Ti(5)-O(5)    &  1.849  &  1.842   \\
    Ti(2)-O(2) & 1.973    & 1.974  & Ti(5)-O(6)    &  2.055  &  2.053   \\ 
    Ti(2)-O(5) & 2.004    & 2.007  & Ti(5)-O(7)    &  2.145  &  2.155  \\
    Ti(3)-O(1) & 1.903    & 1.906  & Ti(5)-O(7)    &  2.002  &  1.992   \\
    Ti(3)-O(2) & 1.939    & 1.929  & Ti(5)-O(8)    &  1.956  &  1.943   \\ 
    Ti(3)-O(3) & 1.991    & 1.976  & Ti(5)-O(9)    &  2.045  &  2.040  \\ 
    Ti(3)-O(6) & 2.029    & 2.035  & Ti(6)-O(4)    &  1.879  &  1.883   \\
    Ti(3)-O(7) & 2.075    & 2.071  & Ti(6)-O(6)    &  2.143  &  2.133   \\ 
    Ti(3)-O(8) & 2.081    & 2.086  & Ti(6)-O(7)    &  1.991  &  2.003  \\
    Ti(4)-O(1) & 1.942    & 1.943  & Ti(6)-O(8)    &  1.855  &  1.856   \\
    Ti(4)-O(3) & 1.902    & 1.907  & Ti(6)-O(9)    &  2.196  &  2.186   \\ 
    Ti(4)-O(4) & 2.007    & 1.995  & Ti(6)-O(9)    &  1.976  &  1.981  \\
%\hline
    \end{tabular}
\end{ruledtabular}
\end{table}

\begin{table}
 \caption{Ti-Ti interatomic distances (in \AA) in the LT (115~K), IT (135~K) and HT( 295~K) phases of Ti$_5$O$_9$ according to structural parameters from Ref.~\onlinecite{Marezio_1977} }\label{tab:table_3}
\begin{ruledtabular}
\begin{tabular}{c c c c }
      & 115~K & 135~K & 295~K \\
  \hline
  Ti(1)-Ti(3) & 2.930 & 2.932 & 2.928 \\
  Ti(2)-Ti(4) & 2.906 & 2.908 & 2.924 \\
  Ti(3)-Ti(5) & 2.979 & 2.989 & 3.004 \\
  Ti(4)-Ti(6) & 3.037 & 3.043 & 3.041 \\
  Ti(5)-Ti(6) & 2.827 & 2.816 & 2.829 \\
 % \hline
    \end{tabular}
\end{ruledtabular}
\end{table}

As mentioned earlier, Ti$_5$O$_9$ crystallizes in the triclinic $P\overline{1}$ structure with two formula units per unit cell. There are six crystallographically independent Ti atoms in its unit cell, surrounded by distorted oxygen octahedra (see Fig.\ref{fig:Fig_1} and Tab.\ref{tab:table_2}). Among them Ti(1) and Ti(2) are at the center of symmetry, while the other eight atoms are related by inversion symmetry, namely Ti(3)$\equiv$Ti(7), Ti(4)$\equiv$Ti(8), Ti(5)$\equiv$Ti(9) and Ti(6)$\equiv$Ti(10). These atoms form two crystallographycally independent pseudorutile chains: Ti(5)$-$Ti(3)$-$Ti(1)$-$Ti(3)$-$Ti(5) and Ti(6)$-$Ti(4)$-$Ti(2)$-$Ti(4)$-$Ti(6) \cite{Marezio_1974}. The Ti(5) and Ti(6) atoms are the nearest to the shear planes and their environment is not rutile-, but sesquioxide-like, similarly to the Ti$_4$O$_7$. The Ti(5)-Ti(6) distance is $\sim$2.83~\AA\ and is the shortest in the lattice (see Table \ref{tab:table_3}). The Oxygen octahedra around these atoms are severely distorted (the difference in Ti-O distances is $\sim$0.3~\AA\ ) (see Table \ref{tab:table_2}). According to Ref.~\onlinecite{Marezio_1977}, the Ti(5)$-$Ti(3)$-$Ti(1)$-$Ti(3)$-$Ti(5) chains are more distorted than the Ti(6)$-$Ti(4)$-$Ti(2)$-$Ti(4)$-$Ti(6) one, and thus have a slightly larger separation of charge. However, as can be seen from Tab.\ref{tab:table_3}, these distortions are very small and structurally the low-temperature phases are almost identical to the HT phase. 
This is not the case for Ti$_4$O$_7$, where the formation of bipolarons in the LT phase and their rearrangement into polarons and bipolarons in the IT phase with the consequent collapse in the HT phase are clearly related with the electron-lattice coupling \cite{}. The transition from HT to LT phase in Ti$_4$O$_7$ is accompanied by an essential decrease in the Ti-Ti distances ($\sim$0.2~\AA) with a consequent charge localization at Ti ions, involved in longer Ti-O and shorter Ti-Ti bonds. Interestingly, in the isostructural V$_4$O$_7$ no changes in cation-cation distances larger than $\sim$0.06~\AA\ are observed at the phase transition, despite the clear separation of charge into $V^{3+}$ and $V^{4+}$ in the alternate chains. Similar behavior applies to another member of the vanadium series, V$_5$O$_9$, which undergoes a metal-insulator transition at 135~K \cite{Marezio_1974}. Although incomplete, V-V dimerization is clearly observed in V$_4$O$_7$ \cite{Botana_2011}, while it is negligible in V$_5$O$_9$ \cite{Obara_2010}. It was speculated that there are many different bonding patterns in the latter two compounds, and that an average of these patterns is observed by the classical x-ray diffraction methods \cite{Marezio_1974}.
\begin{figure}[htbp]
	\begin{center}
		\includegraphics[width=1.0\hsize]{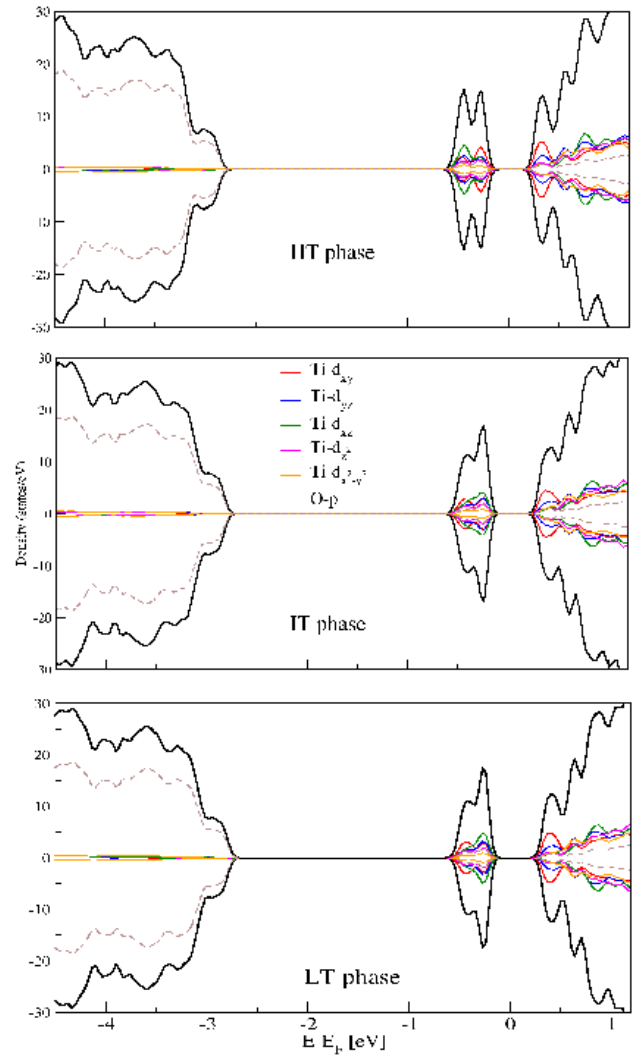}
             \caption{(color online) The GGA+$U$ spin-polarized density of electronic states (DOS), calculated for the LT, IT and HT phases of Ti$_5$O$_9$ at $U=5$ for the AFM spin allignment (the lowest-energy configuration on Fig.\ref{fig:Fig_5}). Positive (negative) values indicate spin-up (spin-down) DOS. \label{fig:Fig_4}}
	\end{center}
\end{figure}

Obviously, if the insulating behavior was induced by the lattice distortions only, the crystalline symmetry below the phase transition would be lower than in the metallic region. However, this is not the case in the Ti$_5$O$_9$ crystal, although its LT phase is slightly more distorted than the HT one. 
To find out how these small lattice distortions at the phase transitions affect the electronic properties of Ti$_5$O$_9$, we performed total-energy calculations for the experimental structures \cite{Marezio_1977} of all three phases. 
Figure \ref{fig:Fig_4} shows the density of states (DOS) of Ti$_5$O$_9$, calculated at $U$=5~eV for the LT, IT and HT structures (as reported by Marezio \textit{et al} \cite{Marezio_1977}), assuming an AFM spin ordering with zero total moment per cell. One can clearly see that the DOS for all the three phases is very similar to each other. They all are insulating with the band gap in the LT phase of about 0.43~eV and only slightly smaller gap of 0.38~eV in the IT and HT phases. The corresponding charge distribution is similar from phase to phase with only a slight difference in the Ti magnetic moments (see Tables~\ref{tab:table_4} and \ref{tab:table_5} for $M_{\mathrm{tot}}=0$). Therefore, although the peculiarities of the crystalline structure of Ti$_5$O$_9$ are certainly reflected in the rich variety of the possible charge order paterns, the lattice distortions at the phase transitions are rather insignificant to have a crucial effect on the conducting properties of the material.  

\subsection{Magnetic and charge order}
\label{sec:Magnetic and charge order}

To find the ground state magnetic ordering and which Ti atoms participate in it, we have considered an antiferromagnetic and a ferromagnetic (FM) spin configuration along with a non-magnetic one (which turned out to be much higher in energy for all three phases). First, we have considered the effect of electron correlations on the magnetic behavior of the crystal. For the experimental structures we have found that at $U\leq4$~eV the LT phase is FM and metallic, while for larger $U$ it tends to be an AFM insulator (see Table~\ref{tab:table_1}). Depending on the chosen $U$ and the resulting spin configuration, the calculated magnetic moments on each of the inequivalent Ti atoms vary within the range 0.1-0.8 $\mu_B$. This is in contrast to Ti$_4$O$_7$, for which the ionic magnetic moments in the LT phase were calculated to be robust with respect to their relative orientation and thus were linked to the local structure around Ti \cite{Weissmann_2011}.  
As follows from Table \ref{tab:table_1}, even in the FM state the magnetic moment is distributed among all the Ti atoms somewhat inhomogeneously. In the AFM state the charge disproportination is more pronounced within the whole range of $U$ values and non-zero moment is observed only at Ti(3), Ti(4) and Ti(5) sites. While Ti(4) and Ti(5) magnetic moments increase with increasing $U$, the moment at Ti(3) site starts to decrease for $U>4$ and is completely suppressed at $U=6$~eV (nominal Ti$^{3+}$ and Ti$^{4+}$ ions become clearly distinguishable). When Ti(3) moment becomes zero, the charge order pattern is symmetric with respect to the inversion. At the FM spin allignment the behavior is somewhat different: the charge order pattern is less symmetric, but more homogeneous, and it is less sensitive to the changes of $U$ value. As a result, the energy gap is zero for the whole range of $U$ values.  

\begin{figure}[htbp]
	\begin{center}
		\includegraphics[width=0.6\hsize]{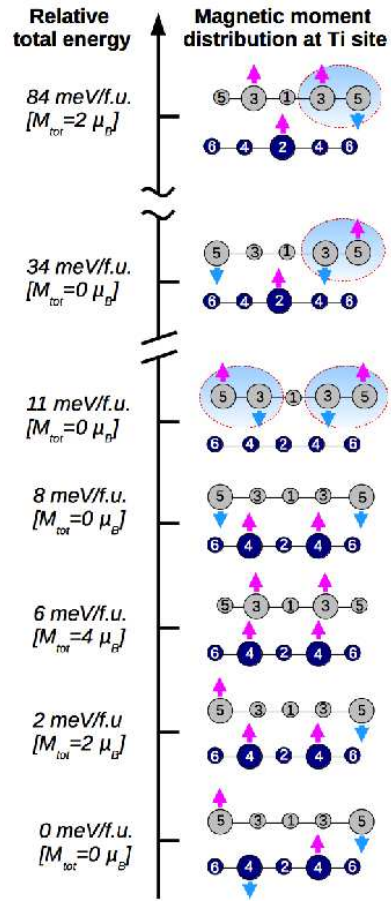}
 		\caption{(color online) A schematic diagram showing the relative stability of several lowest-energy charge and magnetically ordered states, obtained for the LT phase taking into account structural relaxations. Shaded area within a red-dashed oval depicts Ti$^{3+}$-Ti$^{3+}$ singlet pairs (bipolarons). Dark blue and grey colors are used to distinguish between Ti in adjacent pseudorutile chains. Big (small) balls denote Ti$^{3+}$ (Ti$^{4+}$) ions. \label{fig:Fig_5}}
	\end{center}
\end{figure}

We have performed fixed spin moment calculations for all three phases by calculating the total energy at the constrained magnetic moment of the unit cell. The results, obtained for $U=5$~eV, are presented in Tab.\ref{tab:table_4} and Tab.\ref{tab:table_5}. It is clear from the listed data that there are several stable magnetic configurations with a very small relative energy difference, namely the configurations with total moment of 0 (AFM), 2 and 4~$\mu_B$/cell. Moreover, for each of the magnetic configurations we have obtained a whole set of different possible charge oder patterns (see Fig.\ref{fig:Fig_5}); therefore, we present here only those which in our calculations have the lowest energy. 

\subsubsection{Low-temperature phase}

For the experimental structure of the LT phase the configuration with the total magnetic moment $M_{\mathrm{tot}}=2~\mu_B$/cell is practically degenerate with the AFM configuration with $M_{\mathrm{tot}}=0$ (the energy difference is only 6 meV/f.u.) (see Table \ref{tab:table_4}). The FM configuration with $M_{\mathrm{tot}}=4~\mu_B$/cell is 20 meV/f.u. higher in energy than the AFM one. We observe stronger localization of the charge, and hence, larger band gap, for the configurations where more Ti moments are alligned antiparallel to each other: the band gap is the largest for the AFM state and it decreases to zero for the FM one. Therefore, opening of the gap in Ti$_5$O$_9$ when going from high to low temperatures might be closely related with the AFM ordering of the unpaired spins. 
This is in line with the fact that experimentally the N\'eel temperature and the semiconductor-metal transition temperatures in this material nearly coincide. %\textit{Note also the inversion symmetry breaking in the charge distribution for the configuration with $M_{\mathrm{tot}}=2~\mu_B$/cell.} 

\begin{table}
 \caption{Magnetic moments on Ti atoms (in $\mu_B$), total energies $\Delta{E}$ with respect to the lowest-energy magnetic configuration (in meV/f.u.) and the band gap $E_g$ (in eV), calculated for the experimental and relaxed (theoretical) structures of LT phase.}\label{tab:table_4}
\begin{ruledtabular}
   \begin{tabular}{c|r@{ }lc|r@{ }lc|r@{ }lc}   
    M$_{\mathrm{tot}}$ & \multicolumn{3}{c|}{$0~\mu_B$/cell}   & \multicolumn{3}{c|}{$2~\mu_B$/cell}  & \multicolumn{3}{c}{$4~\mu_B$/cell}\\
    \hline
  
$\mu_{\mathrm{Ti}}$ & \multicolumn{2}{c}{exp.} & relaxed & \multicolumn{2}{c}{exp.} &  relaxed & \multicolumn{2}{c}{exp.} &  relaxed \\ 
    \hline
    Ti(1)&            0      & & 0       &  0       & & 0       & 0.14 & & 0 \\
    Ti(2)&            0      & & 0       &  0.31    & & 0       & 0.54 & & 0  \\ 
    Ti(3)&            $-0.28$& & 0       &  $-0.17$ & & 0       & 0.44 & & 0  \\ 
    Ti(4)&            0.71   & & 0.83    &  0.68    & & 0.84    & 0.37 & & 0.85 \\
    Ti(5)&            $-0.53$& & $-0.81$ &  $-0.56$ & & $-0.80$ & 0.61 & & 0.84  \\
    Ti(6)&            0      & & 0       &  0       & & 0       & 0.17 & & 0     \\
    Ti(7)&            $0.28$ & & 0       &  0.32    & & 0       & 0.44 & & 0     \\ 
    Ti(8)&            $-0.71$& & $-0.83$ &  0.67    & & 0.84    & 0.37 & & 0.85  \\
    Ti(9)&            $0.53$ & & $0.81$  &  0.55    & & 0.83    & 0.61 & & 0.84  \\
   Ti(10)&            0      & & 0       &  0       & & 0       & 0.17 & & 0    \\
\hline
    $\Delta{E}$ &     6      & & 0       &  0       & & 2       & 26   & & 7 \\
\hline
    $E_g$       &     0.46   & & 0.9     &  0.19    & & 0.82    & 0    & & 0.67   \\
    \end{tabular}
\end{ruledtabular}
\end{table}

Interestingly, as follows from Table \ref{tab:table_4}, when structural relaxations are taken into account, the lowest energy charge distribution is the same for all the three magnetic configurations. It is such that Ti$^{3+}$ ions (polarons) are at the position of Ti(4) and Ti(5) cations. Within a pseudorutile chain they are well separated by either one or three Ti$^{4+}$ ions (Fig.\ref{fig:Fig_3}), retaining the inversion symmetry. While Ti$^{3+}$ in the Ti(5) position has a common face with the Ti$^{4+}$ at Ti(6) site across the shear plane, the Ti$^{3+}$ in the Ti(4) position is close to the middle of the pseudorutile chain. It is clear from Fig.~\ref{fig:Fig_3}(b) that there is no direct overlap between the $t_{2g}$ orbitals at Ti(4) and Ti(5) sites
and the interaction between the corresponding electrons is realized through the hopping via bridging oxygens. 

Several decades ago, James and Catlow \cite{James_1977} theoretically predicted that the electrons created by the reduction of rutile or higher Magn\'eli phases should be trapped by the shear plane. In Ti$_5$O$_9$ this would result in charge-order state with Ti$^{3+}$ ions occupying Ti(5) and Ti(6) sites. However, in our calculations such state is $\sim$148 meV/f.u. less stable than the lowest-energy one with Ti$^{3+}$ at Ti(5) and Ti(4) sites. This means that not all electrons occupy the sites adjacent to the shear planes, but prefer the localization at the nearest neighbor sites.

In contrast to the experimental structures, the localization of the charges in the relaxed structures is much stronger and it results in the insulating behavior even for the FM spin allignment. 
The total energy difference between the states with $M_{\mathrm{tot}}=0$, 2 and $4~\mu_B$/cell is less than 10 meV/f.u.
% and it defines the energy cost of flipping a single spin at Ti sites without changing the charge order. 
The fact that it is so small indicates that the charge and spin order is unstable. A large separation between the Ti$^{3+}$ ions results in rather weak magnetic interactions, which is reflected in the small energy difference between the states with the same charge distribution, but different number of parallel/antiparallel spins.
This is in line with the conclusions made by Danley and Muley \cite{Danley_1972}, who observed small Weiss constant and hence weak interaction between the unpaired spins in Ti$_5$O$_9$. This is different from Ti$_4$O$_7$, for which the AFM interaction between the spins within the Ti$^{3+}$-Ti$^{3+}$ pairs was found to be rather strong \cite{Leonov_2006}, although the inter-pair coupling was calculated to be much weaker. We have also considered the states, where four unpaired spins at Ti sites are distributed in the way to form bipolarons within the same pseudorutile chain, as well as the state with coexisting bipolarons and polarons. These states were obtained to be 11~meV/f.u. and 34~meV/f.u., correspondingly, higher in energy than the ground state and are both insulating. Therefore, the ground state of the LT phase of Ti$_5$O$_9$ is semiconducting,
however its charge order pattern is not unique.

\subsubsection{Intermediate- and high-temperature phases}

Since our \textit{ab initio} calculations are performed at $T=0$~K temperature, for the IT and HT phases we have considered only the experimental structures as reported by Marezio \textit{et al} \cite{Marezio_1977}. 
In analogy to the LT phase, we observe several stable magnetic configurations for the IT and HT phases, which are almost degenerate within our calculation error: these are the configurations with the total magnetic moment of 0, 2 and 4~$\mu_{\mathrm{B}}$/cell. For each of these magnetic configurations our fixed-spin moment calculations revealed several possible charge order patterns, but in Table~\ref{tab:table_5} we present only the lowest-energy ones. For the IT phase the calculated band gap is the largest for the configuration with $M_{\mathrm{tot}}=0~\mu_B$/cell (about 0.4 eV) and it is almost the same as the gap calculated for the experimental structure of the LT phase. The gap is somewhat smaller for the configuration with $M_{\mathrm{tot}}=2~\mu_B$/cell and completely vanishes when the spins are alligned ferromagnetically due to more homogeneous distribution of the valence charge. Similar tendency is also observed for the HT phase with the only difference being that the gap (although very small) is non-zero for the FM configuration. It should be noted that for the HT phase there is another FM state, which is metallic and has a more homogeneous charge distribution, but this state is about 30~meV/f.u. higher in energy.  

The magnetic moments do not vary much from phase to phase, which once again indicates a secondary role of the crystal structure in the electronic properties of the Ti$_5$O$_9$. The charge distribution for the considered configurations resembles the one observed in the LT phase, however, it is slightly more localized for the FM configuration in the HT phase than in the LT or IT phases. However, the magnetic moment distribution remains inhomgeneous even for the FM configuration in the HT phase and does not correspond to the suggested Ti$^{3.5}$ picture. Therefore, the local environment and distortions play a visible role in the magnetic moment distribution in this material.

\begin{table}
 \caption{Magnetic moments on Ti atoms (in $\mu_B$), $\Delta{E}$ total energies with respect to the lowest-energy magnetic configuration (in meV/f.u.) and the band gap $E_g$ (in eV), calculated for the experimental structures of the IT and HT phases.}\label{tab:table_5}
\begin{ruledtabular}
   \begin{tabular}{c|r@{ }lc|r@{ }lc|r@{ }lc}   
    M$_{\mathrm{tot}}$ & \multicolumn{3}{c|}{$0~\mu_B$/cell}   & \multicolumn{3}{c|}{$2~\mu_B$/cell}  & \multicolumn{3}{c}{$4~\mu_B$/cell}\\
    \hline
  
$\mu_{\mathrm{Ti}}$ & \multicolumn{2}{c}{IT} & HT & \multicolumn{2}{c}{IT} &  HT & \multicolumn{2}{c}{IT} &  HT \\ 
    \hline
    Ti(1)&            0      & & 0       &  0       & & 0       & 0.20 & & 0 \\
    Ti(2)&            0      & & 0       &  0.35    & & 0.27    & 0.68 & & 0.29  \\ 
    Ti(3)&            $-0.25$& & $-0.33$ &  $-0.13$ & & $-0.14$ & 0.43 & & 0.41  \\ 
    Ti(4)&            0.71   & & 0.73    &  0.67    & & 0.70    & 0.19 & & 0.70 \\
    Ti(5)&            $-0.55$& & $-0.46$ &  $-0.58$ & & $-0.56$ & 0.69 & & 0.47  \\
    Ti(6)&            0      & & 0       &  0       & & 0       & 0.19 & & 0.14     \\
    Ti(7)&            $0.25$ & & 0.33    &  0.29    & & 0.46    & 0.42 & & 0.40     \\ 
    Ti(8)&            $-0.71$& & $-0.73$ &  0.64    & & 0.69    & 0.19 & & 0.70  \\
    Ti(9)&            $0.55$ & & $0.46$  &  0.57    & & 0.39    & 0.69 & & 0.47  \\
   Ti(10)&            0      & & 0       &  0       & & 0       & 0.19 & & 0.14   \\
\hline
    $\Delta{E}$ &     9      & & 8       &  0       & & 2       & 2   & & 0 \\
\hline
    $E_g$       &     0.43   & & 0.38     &  0.15    & & 0.25    & 0    & & 0.04   \\
    \end{tabular}
\end{ruledtabular}
\end{table}

\section{conclusions}
\label{sec:conclusions}

In summary, we have performed first-principles calculations to study the electronic and magnetic propeties of the Ti$_5$O$_9$ crystal, which belongs to the homologous series of Ti$_n$O$_{2n-1}$ oxides, known as Magn\'eli phases. From our results it is difficult to conclude on which charge distribution is realized in this material at different temperatures, since for all three phases we find several quasidegenerate magnetic solutions. We therefore assume that the charge order within the cation sublattice in Ti$_5$O$_9$ is not unique and thus unstable, which makes its experimental detection at finite temperatures difficult. We suggest that in contrast to Ti$_4$O$_7$, the formation of Ti$^{3+}$-Ti$^{3+}$ pairs in this compound is less likely, although not completely impossible, since a bipolaronic state is only slightly higher in energy than the ground state. Our calculations perfomed for all three phases showed that crystal structure changes at the transitions make a minor influence on the electronic structure of the compounds.
Although the electronic correlations are important for the stabilization of the insulating ground state, the phase transitions in this compound are related to a greater extent to a reordering of the unpaired spins. This conclusion agrees well with the antiferromagnetism model presented by Adler in Ref.\onlinecite{Adler_1968}, where the author considered different scenarios for the metal-nonmetal transitions in transition metal oxides and sulfides.  
Thus, the mechanism behind the phase transitions in Ti$_5$O$_9$ must be a complex interrelation between the electronic correlations, electron-lattice, as well spin-lattice coupling.

\label{sec:acknowledgement}
We appreciate the support of the J\"ulich Supercomputing Centre (Project JIFF38) and gratefully acknowledge the support of the SFB917-Nanoswitches and Young Investigators Group Program of the Helmholtz Association, Contract VH-NG-409. 

\label{sec:bibliography}

\end{document}